\documentclass[usegraphicx,usenatbib]{mn2e}

\usepackage{graphicx}
\usepackage{natbib}
\usepackage{fixltx2e}
\usepackage{mathptmx}
\usepackage{amsmath}
\usepackage{url}
\usepackage{times}
\usepackage{captcont}
\usepackage{epsfig}
\usepackage{epstopdf}
\usepackage{ulem}

\newcommand{\Mpc}{\rm\; Mpc}
\newcommand{\kpc}{\rm\; kpc}

\newcommand{\km}{\rm\; km}

\newcommand{\hr}{\rm\; hr}
\newcommand{\yr}{\rm\; yr}
\newcommand{\Gyr}{\rm\; Gyr}

\newcommand{\s}{\rm\; s}

\newcommand{\ks}{\rm\; ks}

\newcommand{\GHz}{\rm\; GHz}
\newcommand{\MHz}{\rm\; MHz}

\newcommand{\keV}{\rm\; keV}

\newcommand{\erg}{\rm\; erg}

\newcommand{\mJy}{\rm\; mJy}

\newcommand{\ergps}{\hbox{$\erg\s^{-1}\,$}}

\newcommand{\kmps}{\hbox{$\km\s^{-1}\,$}}

\newcommand{\kmpspMpc}{\hbox{$\kmps\Mpc^{-1}\,$}}

\newcommand{\amin}{\rm\; arcmin}

\newcommand{\asec}{\rm\; arcsec}

\begin{document}

\title[The merging cluster Abell 2146]{A merger mystery: no extended radio emission in the merging cluster Abell 2146}
\author[H.R. Russell et al.]
       {\parbox[]{7.in}{H.~R. Russell$^1$\thanks{E-mail:
             helen.russell@uwaterloo.ca}, R.~J. van Weeren$^2$,
           A.~C. Edge$^3$, B.~R. McNamara$^{1,4,5}$,
           J.~S. Sanders$^6$, A.~C. Fabian$^6$, S.~A. Baum$^7$, R.~E.~A. Canning$^6$, M. Donahue$^8$ and C.~P. O'Dea$^9$\\
           \footnotesize 
           $^1$ Department of Physics and Astronomy, University of Waterloo, Waterloo, ON N2L 3G1, Canada\\
           $^2$ Leiden Observatory, Leiden University, P.O. Box 9513, NL-2300 RA Leiden, The Netherlands\\
           $^3$ Department of Physics, Durham University, Durham DH1 3LE\\
           $^4$ Perimeter Institute for Theoretical Physics, Waterloo, Canada\\
           $^5$ Harvard-Smithsonian Center for Astrophysics, 60 Garden Street, Cambridge, MA 02138, USA\\
           $^6$ Institute of Astronomy, Madingley Road, Cambridge CB3 0HA\\
           $^7$ Center for Imaging Science, Rochester Institute of Technology, Rochester, NY 14623, USA\\
           $^8$ Department of Physics and Astronomy, Michigan State University, East Lansing, MI 48824, USA\\
           $^9$ Department of Physics, Rochester Institute of Technology,
    Rochester, NY 14623, USA
  }
}

\maketitle

\begin{abstract}
We present a new $400\ks$ \textit{Chandra} X-ray observation and a
\textit{GMRT} radio observation at $325\MHz$ of the merging galaxy
cluster Abell 2146.  The \textit{Chandra} observation reveals detailed
structure associated with the major merger event including the Mach
$M=2.1\pm0.2$ bow shock located ahead of the dense subcluster core and
the first known example of an upstream shock ($M=1.6\pm0.1$).
Surprisingly, the deep \textit{GMRT} observation at $325\MHz$ does not
detect any extended radio emission associated with either shock front.
All other merging galaxy clusters with X-ray detected shock fronts,
including the Bullet cluster, Abell 520, Abell 754 and Abell 2744, and
clusters with candidate shock fronts have detected radio relics or
radio halo edges coincident with the shocks.  We consider several
possible factors which could affect the formation of radio relics,
including the shock strength and the presence of a pre-existing
electron population, but do not find a favourable explanation for this
result.  We calculate a $3\sigma$ upper limit of $13\mJy$ on extended
radio emission, which is significantly below the radio power expected
by the observed $P_{\mathrm{radio}}-L_{\mathrm{X}}$ correlation
for merging systems.  The lack of an extended radio halo in Abell 2146
maybe due to the low cluster mass relative to the majority of merging
galaxy clusters with detected radio halos.
\end{abstract}


\section{Introduction}
\label{sec:intro}

\begin{figure*}
\begin{minipage}{\textwidth}
\centering
\raisebox{0.07\height}{\includegraphics[width=0.33\columnwidth]{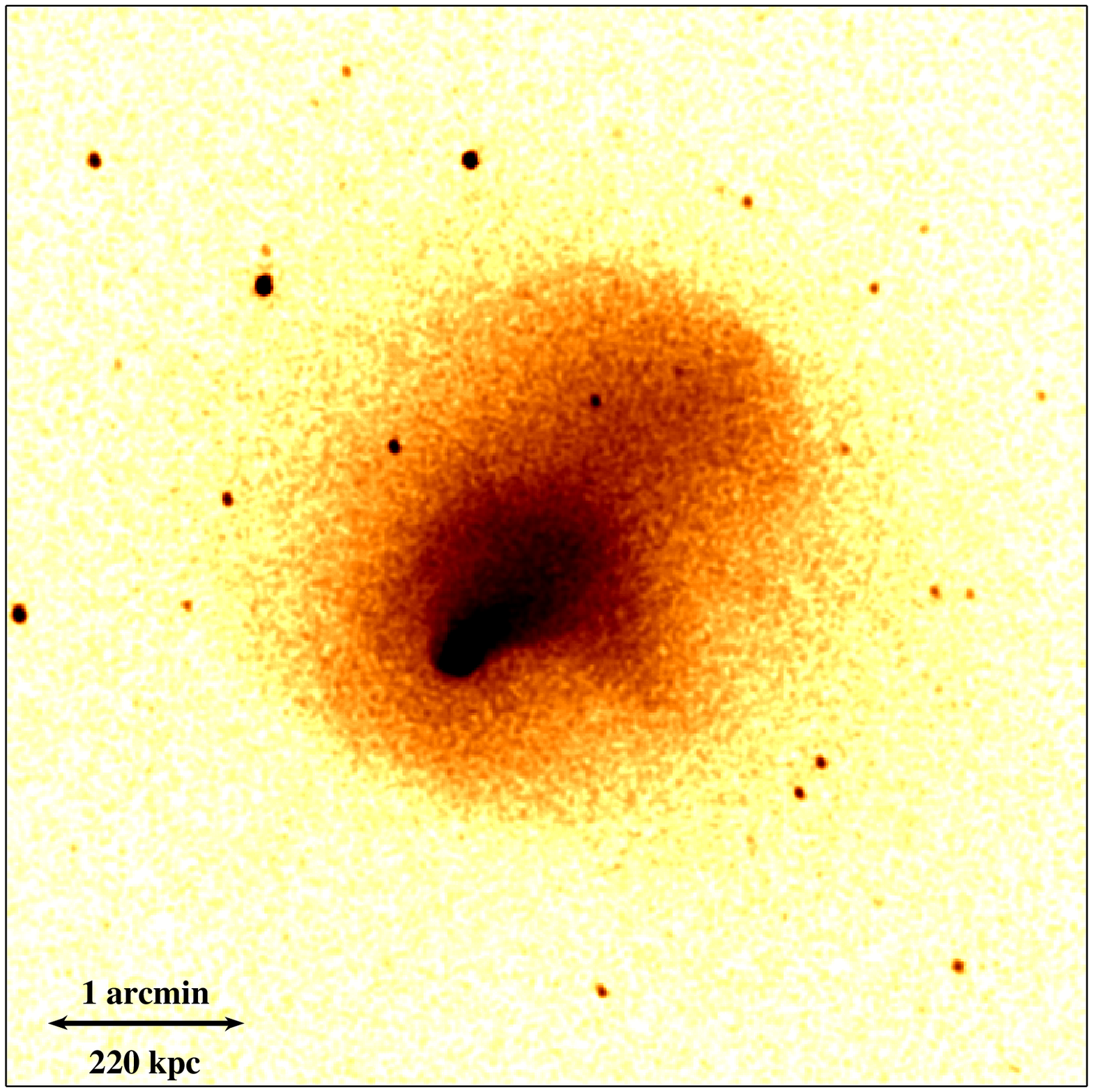}}
\includegraphics[width=0.33\columnwidth]{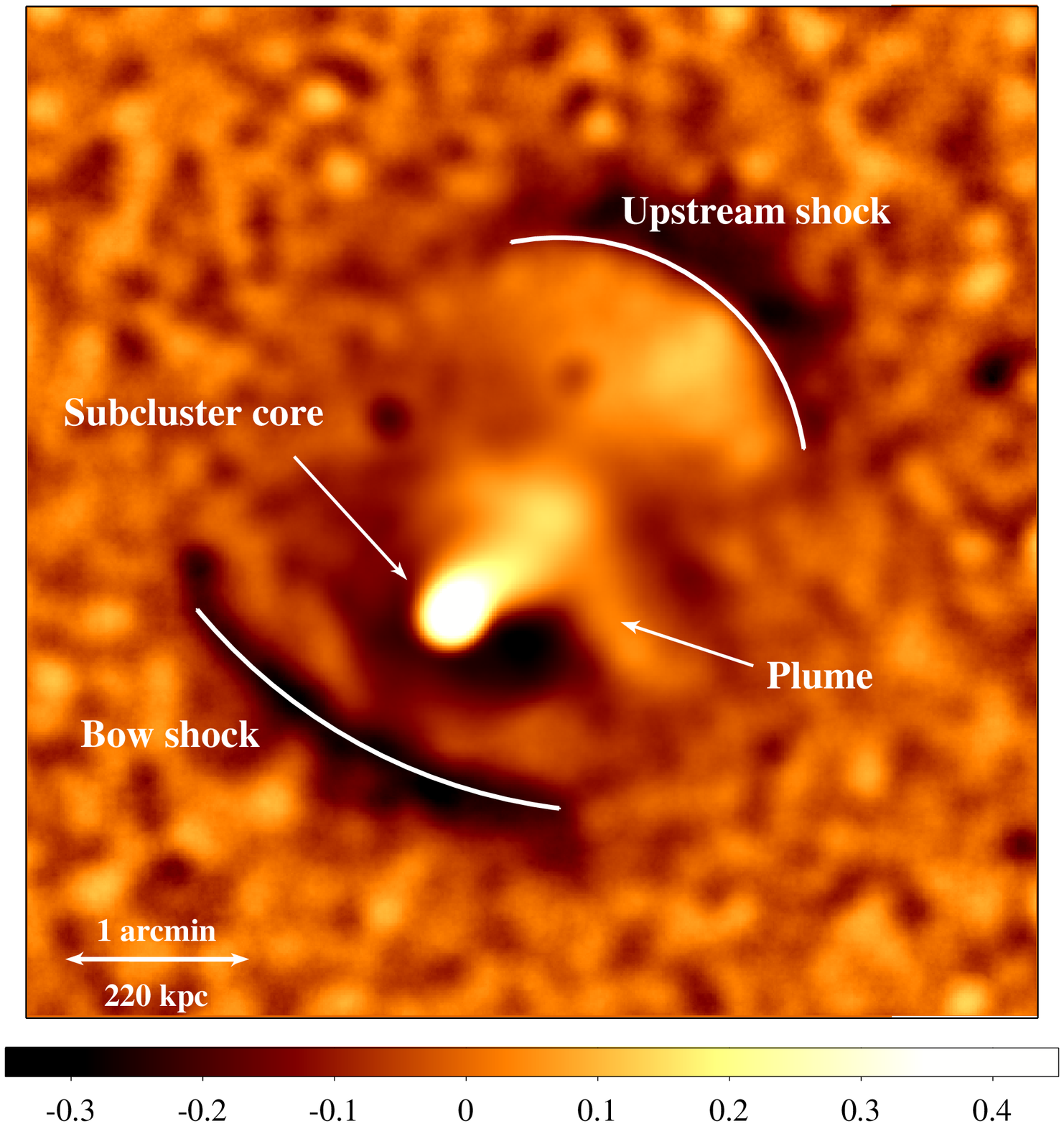}
\raisebox{0.07\height}{\includegraphics[width=0.33\columnwidth]{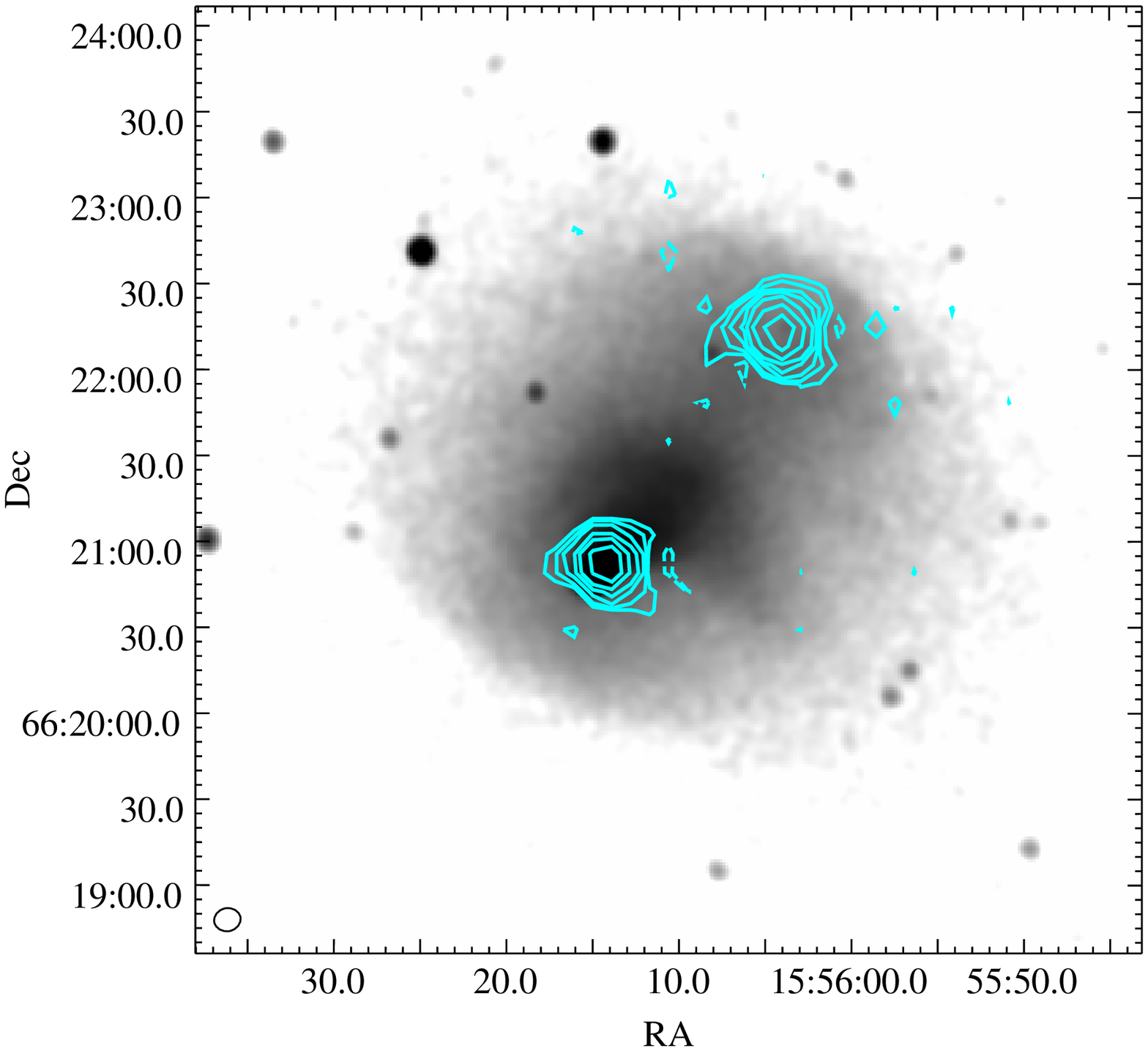}}
\vspace{-2\baselineskip}
\caption{Left: Exposure-corrected image in the 0.3--7.0$\keV$ energy
  band smoothed with a 2D Gaussian $\sigma=1.5\asec$ (North is up and
  East is to the left).  Centre: Unsharp-masked image created by
  subtracting images smoothed by 2D Gaussians with $\sigma=5$ and
  $20\asec$ and dividing by the sum of the two images.  Point sources
  were removed before unsharp-masking.  Note that a point source has
  been removed at the NE end of the bow shock.  Both left and centre
  images are $5.5\times5.5\amin$.  Right: \textit{Chandra} image
  smoothed with a 2D Gaussian $\sigma=1.5\asec$.  \textit{GMRT}
  $325\MHz$ contours are superimposed with levels drawn at
  $[-1,1,2,4,8,\ldots] \times 4\sigma_{rms}$.  Negative contours are
  shown by the dotted lines.  The beam size is $9.3\asec \times
  8.1\asec$ and is shown in the bottom left corner.}
\label{fig:mainimages}
\end{minipage}
\end{figure*}

Galaxy clusters are formed through gravitational infall and mergers of
smaller subclusters, which collide at velocities of several
thousand $\km\s^{-1}$.  The total kinetic energy of these mergers can
reach $\geq10^{64}\erg$, a significant fraction of which is dissipated
by shocks and turbulence over the merger lifetime (for a review see
\citealt{Markevitch07}).  Shocks and turbulence generated by the
merger are also expected to amplify cluster magnetic fields and
accelerate relativistic particles.  These non-thermal phenomena have
been revealed through the detection of Mpc-scale synchrotron radio
halos (for recent reviews see \citealt{Feretti08};
\citealt{Ferrari08}) and inverse-Compton hard X-ray emission
(eg. \citealt{Fusco-Femiano05} but see also \citealt{Wik09}).  Major
mergers can also produce observable separations between the
intracluster medium (ICM), which represents the bulk of the baryons,
and the dark matter halo.  The combination of X-ray and gravitational
lensing studies of these clusters has produced the most striking
evidence for the existence of dark matter
(eg. \citealt{Clowe04}).

Shock fronts provide a key observational tool in the study of merging
systems.  They can be used to determine the velocity and kinematics of
the merger and to study the conditions and transport processes in the
ICM (eg. \citealt{Markevitch07}).  Shock fronts are detected as sharp
discontinuities in X-ray surface brightness maps and, with spatially
resolved spectroscopy, provide a method for measuring the gas
velocities in the plane of the sky.  While many clusters are found to
have shock-heated regions (eg. \citealt{Henry95};
\citealt{Markevitch96}), the detection of a sharp density edge and an
unambiguous temperature jump is rare and only a handful of shock fronts are
currently known (the Bullet cluster, \citealt{Markevitch02}; Abell
520, \citealt{Markevitch05}; two in Abell 2146, \citealt{Russell10};
Abell 754, \citealt{Macario11} and Abell 2744, \citealt{Owers11}).

All merging galaxy clusters with X-ray detected shock fronts also
contain large, diffuse radio halos covering $\sim1\Mpc$ in size
(eg. \citealt{Feretti08}).  Due to synchrotron and inverse-Compton
losses, the relativistic electrons have relatively short lifetimes,
only $10^7-10^8\yr$, which is significantly shorter than the diffusion
time necessary to cover their observed extent.  Therefore, a mechanism
is required that produces local particle acceleration over the whole
cluster volume (\citealt{Jaffe77}).  The possibilities include
`primary' models, where a pre-existing electron population is
reaccelerated by turbulence or shocks (eg. \citealt{Brunetti01};
\citealt{Petrosian01}), and `secondary' models, which rely on the
continuous injection of relativistic electrons by hadronic collisions
between thermal ions and cosmic rays (eg. \citealt{Dennison80}).  

The observed correlations between extended radio emission and cluster
mergers favour electron reacceleration models (eg. \citealt{Buote01};
\citealt{Feretti04}; \citealt{Feretti08}).  Recent observations
suggest that radio halos are likely generated by turbulent processes
induced by a merger (eg. \citealt{Brunetti08}; \citealt{Dolag08}) and
radio relics, which are elongated structures located in the cluster
outskirts, may be produced by Fermi-I diffusive shock acceleration of
ICM electrons (\citealt{Ensslin98}).  \citet{Keshet10} proposed a
hadronic model where magnetic fields are amplified by mergers and
produce radio halos. However, recently \citet{Bonafede11} found no
difference in fractional polarization properties between clusters with
and without halos implying there no difference in the magnetic fields
between the two.

In this Letter, we present results from a deep \textit{Chandra}
observation of the merging cluster Abell 2146 and a \textit{GMRT}
radio observation at $325\MHz$, which was aimed at detecting any
extended radio emission.  We assume $H_0=70\kmpspMpc$, $\Omega_m=0.3$
and $\Omega_\Lambda=0.7$, translating to a scale of $3.7\kpc$ per
arcsec at the redshift $z=0.234$ of Abell 2146.  All errors are
$1\sigma$ unless otherwise noted.

\section{\textit{Chandra} X-ray observations}
\label{sec:chandradata}
Abell 2146 was observed with the \textit{Chandra} ACIS-I detector for
a total of $377\ks$ split into eight separate observations between
August and October, 2010.  These new observations were combined with
the archival ACIS-S observations, $42.1\ks$, taken in April, 2009
(\citealt{Russell10}).  All datasets were reprocessed following the
standard methods detailed in \citet{Russell10} and using CIAO 4.3 and
CALDB 4.4.0 provided by the \textit{Chandra} X-ray Center.
There were no major flares in any of the observations of Abell 2146
producing a final exposure of $418\ks$.  The cleaned events files were
reprojected to the position of the obs. ID 12245 dataset.

Fig. \ref{fig:mainimages} (left) shows an exposure-corrected X-ray
image produced by combining all of the individual datasets.  The X-ray
morphology reveals a major merger where a subcluster containing a
dense, cool core has recently passed through the centre of the primary
cluster.  The cluster gas is extended along the merger axis from the
NW core collision site to the subcluster which is travelling towards
the SE and disintegrating under ram pressure.  The leading and SE
edges of the subcluster core form a narrow, continuous cold front with
a sharp density jump of a factor of $\sim3$. In contrast, the deep
\textit{Chandra} observations have revealed complex structures in the
tail of stripped gas from the subcluster core, including what appear
to be developing Kelvin-Helmholtz (K-H) instabilities along the NW
edge.  There is also an extended cool plume of emission to the SW from
the subcluster tail, $\sim45\asec$ in length, which appears
perpendicular to the merger axis.

The two shock fronts, reported in \citet{Russell10}, are clearly
visible in the unsharp-masked image (Fig. \ref{fig:mainimages},
centre) as surface brightness edges to the NW and SE.  The SE edge
corresponds to the bow shock, which has formed ahead of the subcluster
core, and can now be traced over $\sim500\kpc$ in length.  The NW edge
corresponds to the upstream shock which has formed in the wake of the
subcluster's passage through the primary cluster core and is
travelling in the opposite direction to the bow shock.  By measuring
the gas density and temperature on either side of each shock front and
applying the Rankine-Hugoniot shock jump conditions
(eg. \citealt{LandauLifshitz59}), we calculated the respective Mach
numbers $M=v/c_s$, where $v$ is the velocity of the preshock gas with
respect to the shock surface and $c_s$ is the velocity of sound in
that gas.  At the bow shock, the density drops by a factor
$\rho_2/\rho_1=2.4\pm0.2$ which gives a Mach number $M=2.1\pm0.2$.
For the upstream shock, the density decreases by
$\rho_2/\rho_1=1.8\pm0.2$ producing $M=1.6\pm0.1$.  The shock
fronts are confirmed by the detection of unambiguous decreases in the
gas temperature coincident with the density drops.  By using the bow
shock velocity and estimating the distance between the subcluster core
and the primary cluster centre, we can calculate the time since the
subcluster passed through the primary cluster core.  For a bow shock
velocity $v=2800^{+400}_{-300}\kmps$ and an estimated distance of
$\sim350\kpc$, the time since core passage is $\sim0.1-0.2\Gyr$.  This
is likely to be an underestimate of the timescale as the subcluster
velocity is presumably lower than the shock velocity
(\citealt{Springel07}).

The deep \textit{Chandra} observation of Abell 2146 presented here
will be analysed in detail in a future paper (Russell et al. in
prep).

\section{\textit{GMRT} observations}
\textit{GMRT} radio continuum observations at $325\MHz$ of Abell 2146
were carried out on August 22, 2010. The \textit{GMRT} software
backend \citep[GSB; ][]{2010ExA....28...25R} was used in spectral line
mode with 512 channels, $4\s$ integration time per visibility and
recording RR and LL polarizations.  The total on source time was
$5.7\hr$. The data were reduced with the NRAO Astronomical Image
Processing System (AIPS) package.  The data were visually inspected
for the presence of radio frequency interference (RFI) and
malfunctioning antennas.  The visibilities were then calibrated for
the bandpass response of the antennas using the calibrator source
3C286.  A few channels at the edge of the band were discarded as they
were too noisy. This was followed by a standard gain calibration using
3C286 and the phase calibrator 1634+627. The flux density of 3C286 was
set according to the \cite{perleyandtaylor} extension to the
\cite{1977A&A....61...99B} scale.

For Abell 2146, a model of the surrounding field was created using the
NVSS survey \citep{1998AJ....115.1693C}. A phase-only self-calibration
against this model was carried out to improve the astrometric
accuracy, followed by three rounds of phase self-calibration
and two final rounds of amplitude and phase self-calibration. To
produce the images, we used the polyhedron method
\citep{1989ASPC....6..259P, 1992A&A...261..353C} to minimize the
effects of non-coplanar baselines. The final image was made using
robust weighting \citep[robust = -1.5,][]{briggs_phd} and manually
placed clean-boxes, giving a beam of $9.3\asec \times 8.1\asec$. The
rms noise in this image measured in a region free of strong
sources is $92\mu$Jy~beam$^{-1}$.

Fig. \ref{fig:mainimages} (right) shows the \textit{Chandra} X-ray
observation overlaid with high resolution \textit{GMRT} $325\MHz$
radio contours.  Two previously known unresolved radio point sources,
which coincide with cluster galaxies (\citealt{Rodriguez11}), were
observed but no extended radio emission was detected.  The NW point
source has a flux of $93\pm9\mJy$ and is associated with a galaxy in
the primary cluster.  The SE point source, flux $47\pm5\mJy$, is
coincident with the subcluster BCG and an X-ray point source detected
in a hard energy band $4-7\keV$.  Using the $325\MHz$ flux and the
NVSS flux at $1.4\GHz$, we determine the spectral index, $\alpha$, for
the NW point source is $\alpha=-0.57\pm0.07$ and for the SE source is
$\alpha=-0.77\pm0.08$, where $F \propto \nu^{\alpha}$.  A lower
resolution \textit{GMRT} image (rms noise $278\mu$Jy~beam$^{-1}$),
with a resolution of $20\asec\times20\asec$, does not show any
additional emission.  Further smoothing did not reveal any extra
large-scale emission.  For a $1\Mpc$ extended source, the $3\sigma$
upper limit at $325\MHz$ was calculated to be $13\mJy$
($3\times(\mathrm{rms\,noise})\sqrt{\mathrm{source\,area}/\mathrm{beam\,area}}$).
The higher resolution image provided a similar limit.  As a further
test of our analysis and flux limit, we inserted simulated radio
halos, modelled by a Gaussian with a FWHM of $500\kpc$, into the
\textit{u-v} data and found that halos $\geq15\mJy$ were detectable
(eg. \citealt{Venturi08}).  Using a conservative estimate of the
spectral index of $\alpha=-1$, we calculate a corresponding upper
limit of $3\mJy$ at $1.4\GHz$.



\section{Discussion}
We present the surprising result that our \textit{GMRT} observation at
$325\MHz$ did not show evidence for any extended radio emission
associated with either shock front in the merging cluster Abell 2146.
A deep \textit{Chandra} observation confirms the strong dynamical
disruption to the cluster cores and the presence of two shock fronts,
each several hundred kpc across and with Mach numbers of $2.1\pm0.2$
and $1.6\pm0.1$.  Observational results have so far suggested a
strong connection between merging clusters hosting shock fronts and
extended radio emission in the form of relics or halos
(eg. \citealt{Giovannini99}; \citealt{GovoniEnsslin01};
\citealt{Feretti04}; \citealt{Venturi07}; \citealt{Giacintucci08};
\citealt{Markevitch10};
\citealt{vanWeeren09,vanWeeren10,vanWeeren11}).  We therefore consider
possible explanations for the lack of extended radio emission in Abell
2146.

\subsection{No radio relics}
All other X-ray detected merger shock fronts, including those in the
Bullet cluster (\citealt{Liang00}; \citealt{Markevitch02}), Abell 520
(\citealt{Govoni01}; \citealt{Markevitch05}), Abell 754
(\citealt{Kassim01}; \citealt{Bacchi03}; \citealt{Macario11}) and
Abell 2744 (\citealt{GovoniEnsslin01}), and shock front candidates are
spatially coincident with radio relics or the edges of radio halos
(eg. \citealt{Markevitch10}).  The two shock fronts in Abell 2146 have
comparable Mach numbers to the other clusters with detected radio
relics or edges, such as Abell 754 $M=1.6$
(\citealt{Macario11}) and Abell 520 $M=2.1$ (\citealt{Markevitch05}).
The lack of radio emission coincident with the shocks is a
surprising result and indicates that other factors can determine the
production of radio relics.

Direct acceleration of cosmic rays by these weak shock fronts is expected to
be an inefficient process (eg. \citealt{Gabici03}; \citealt{Hoeft07}).
Therefore the observations support a scenario where the weak merger
shocks are reaccelerating a pre-existing population of relativistic
electrons (eg. \citealt{Markevitch05}; \citealt{Giacintucci08}) or,
alternatively, re-energising a population of relativistic electrons
trapped in a fossil radio bubble (\citealt{Ensslin01}).  Over the
lifetime of a cluster, there will generally have been a large number
of possible sources of these electrons, such as accretion shocks,
supernovae and AGN.  The pre-existing electron population
could also have accumulated from previous mergers
(eg. \citealt{Sarazinspec99}) or been generated by collisions between
thermal protons and long-lived cosmic rays (\citealt{Dennison80};
\citealt{Blasi99}).

Abell 2146 does not appear to be particularly atypical compared to the
galaxy clusters with detected radio relics.  The global ICM
temperature, $T=6.75\pm0.06\keV$, and luminosity,
$L_{\mathrm{X}}=6.60\pm0.02\times10^{44}\ergps$ ($0.1-2.4\keV$),
indicate a substantial galaxy cluster which is likely to have
experienced significant accretion and previous merger shocks during
its assembly.  Radio-loud AGN have been found generally to be common
in cluster galaxies (eg. \citealt{Best05}).  Our observations of Abell
2146 also confirm the detection of two radio point sources, each
associated with a cluster galaxy.  The SE radio source
(Fig. \ref{fig:mainimages} right) is located in the subcluster BCG and
coincides with a hard X-ray point source, which suggests it is likely
to be AGN.  Although there are no visible substructures or cavities in
the \textit{Chandra} observations indicating current AGN activity, the
subcluster hosts a strong cool core where the gas temperature drops
down to $\sim1\keV$.  Earlier episodes of AGN feedback could have
produced radio bubbles to regulate this cooling.  X-ray spectral
fitting also shows that the ICM has been significantly metal-enriched
by supernova activity (\citealt{Russell10}).  Abell 2146 appears to
contain sufficient sources for producing a pre-existing electron
population.

\subsection{No radio halo}
\begin{figure}
\centering
\includegraphics[width=0.9\columnwidth]{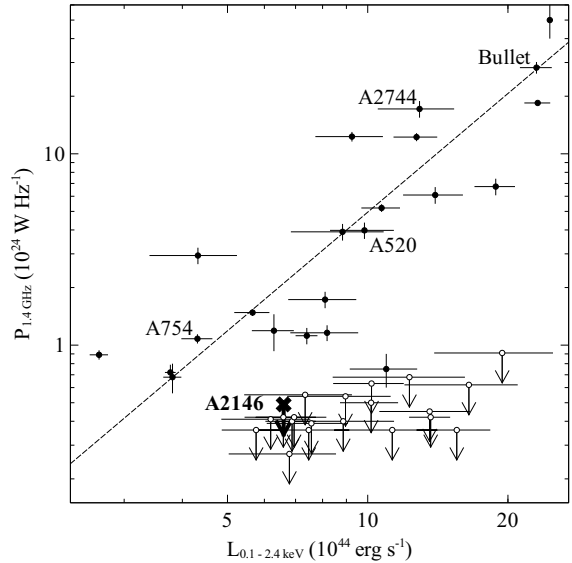}
\vspace{-0.5\baselineskip}
\caption{The $1.4\GHz$ radio halo power plotted against the cluster
  X-ray luminosity in the energy band $0.1-2.4\keV$.  The data points
  and best fit dashed line are from \citet{Brunetti09}.  The
  calculated $3\sigma$ upper limit for Abell 2146 from this work is
  overplotted.  Other clusters with X-ray detected shock fronts are labelled.}
\label{fig:brunettiplot}
\end{figure}

Large radio halos are found exclusively in disturbed and potentially
merging clusters (eg. \citealt{Feretti08}).  Strong correlations exist
between the radio power of halos and the X-ray luminosity and
temperature of their host clusters (eg. \citealt{Colafrancesco99};
\citealt{Liang00}; \citealt{Kempner01}; \citealt{Ensslin02};
\citealt{Cassano06}; \citealt{Giovannini09}).  Recent surveys with the
\textit{GMRT} have found a bi-modal behaviour where radio-halo
clusters and clusters without radio halos are separated in the
$P_{\mathrm{radio}}-L_{\mathrm{X}}$ plane
(Fig. \ref{fig:brunettiplot}; \citealt{Brunetti07}).
\citet{Brunetti09} argue that this bi-modal distribution suggests an
evolutionary cycle where clusters that are undergoing mergers host
radio halos for a period of time.  Later, the clusters become
dynamically relaxed and the relativistic particles cool, producing the
observed upper limits.  Fig. \ref{fig:brunettiplot} shows the
$3\sigma$ upper limit on extended radio emission for Abell
2146 overplotted on the $P_{1.4\GHz}-L_{\mathrm{X}}$ correlation.
This upper limit on Abell 2146 is significantly below the expected
radio power predicted by the observed correlation for merging systems.


The absence of a radio halo in Abell 2146 could be related to the low
cluster mass relative to the majority of the clusters in the GMRT
radio halo sample (\citealt{Venturi07}; \citealt{Brunetti09}).
Observations have found that radio halos are more common in clusters
with higher X-ray luminosity and temperature, which both trace the
mass (eg. \citealt{Giovannini99}; \citealt{Buote01};
\citealt{Cassano06,Cassano08}).  The energy available to accelerate
relativistic particles during a merger is a fraction of the
gravitational potential energy that is released.  Collisions between
more massive galaxy clusters are therefore more likely to produce
radio halos.  The X-ray luminosity and temperature, both relating to
the cluster mass, will also vary over the merger timescale.
Simulations of cluster mergers show that the shocks and compression
produced during core passage can temporarily boost the X-ray
luminosity by a factor of a few (eg. \citealt{Ricker01};
\citealt{Poole07}).  This bias will affect all the merging systems in
Fig. \ref{fig:brunettiplot} to some degree but has the greatest impact
on systems that are closest to core passage, such as Abell 2146.  For
a head-on collision between two subclusters with mass ratio of 3:1
observed only $\sim0.1-0.2\Gyr$ after core passage, the boosts to the
temperature and bolometric X-ray luminosity could produce an increase
of a factor of $\sim2-3$ in the observed $L_{\mathrm{X}}$
(\citealt{Poole07}).  If we `correct' for this factor, the upper limit
for Abell 2146 in Fig. \ref{fig:brunettiplot} is then approximately
consistent with the observed $P_{\mathrm{radio}}-L_{\mathrm{X}}$
correlation.  However, after the maximum around core passage, this
boost to the X-ray luminosity decays rapidly.  For Abell 2146, the
time since core passage has likely been underestimated (section
\ref{sec:chandradata}) and therefore the estimated factor of $\sim2-3$
is an upper limit.  More accurate measurements of the mass of Abell
2146 will be provided by weak-lensing results using \textit{Subaru
  Suprime-Cam} observations (King et al. in prep).

Using the \textit{GMRT} radio halo survey (\citealt{Venturi07}),
\citet{Cassano10} confirmed that radio halos are located
preferentially in dynamically disturbed clusters and identified
several clusters with a disturbed X-ray morphology but no radio
halo. These `outliers' have a relatively low X-ray luminosity
$L_{\mathrm{X}}\leq8\times10^{44}\ergps$, similar to Abell 2146, or
are at a relatively higher redshift, which implies stronger
inverse-Compton losses.  The turbulent reacceleration model for radio
halo formation predicts the existence of very steep spectrum halos
with a spectral cutoff frequency that is dependent on the fraction of
the turbulent energy channelled into electron reacceleration and
therefore the cluster mass (eg. \citealt{Cassano06};
\citealt{Brunetti08}).  Lower frequency
observations, with \textit{LOFAR} for example, could detect radio
halos generated in lower mass mergers, such as these `outlier' clusters
or Abell 2146 (eg. \citealt{CassanoBrunetti10}; \citealt{Venturi11}).

\section{Conclusions}
The deep $400\ks$ \textit{Chandra} observation of Abell 2146 reveals a
major merging event with two shock fronts detected in the X-ray.  The
bow shock, located ahead of the subcluster core, can now be traced to
over $\sim500\kpc$ in length and has a Mach number $M=2.1\pm0.2$.
This cluster also contains the first known example of an upstream
cluster merger shock front, which we determine has a Mach number
$M=1.6\pm0.1$.  Using a new \textit{GMRT} radio observation at
$325\MHz$, we found no evidence for extended radio emission such as a
radio relic associated with either shock or a large radio halo in the
merging cluster Abell 2146.  The absence of radio relics coincident
with the shock fronts is a surprising result.  All other clusters with
unambiguous X-ray detected shock fronts, including the Bullet cluster,
Abell 520, Abell 754 and Abell 2744, and clusters with likely shock
front candidates have radio relics or radio halo edges.  We have
considered several possible factors which could affect the formation
of radio relics, including the shock strength and the presence of a
pre-existing electron population, but do not find a favourable
explanation.  The lack of an extended radio halo in Abell 2146 is
likely due to the low cluster mass relative to the majority of merging
galaxy clusters with detected radio halos.  Using the \textit{GMRT}
radio halo cluster sample, \citet{Cassano10} identified several
clusters with a disturbed X-ray morphology but no radio halo.  The
majority of these clusters have a relatively low X-ray luminosity and
therefore low mass, similar to Abell 2146.  Lower frequency
observations, with \textit{LOFAR} for example, could in the future
detect radio halos generated in lower mass mergers like Abell 2146, which should have ultra-steep radio spectra.

\section*{Acknowledgements}
HRR and BRM acknowledge generous financial support from the Canadian
Space Agency Space Science Enhancement Program.  ACF thanks the Royal
Society for support.  RJvW and REAC acknowledge funding from the Royal
Netherlands Academy of Arts and Sciences and the STFC, respectively.
The \textit{GMRT} is operated by the National Centre for Radio
Astrophysics (NCRA).  The authors are grateful to the \textit{GMRT}
Director and staff for the award of time and executing the
observations.  HRR thanks David Green, Julie Hlavacek-Larrondo,
Roderick Johnstone and Grant Tremblay for helpful discussions.  We
also thank the referee and Gianfranco Brunetti for helpful comments
that improved this letter.

\vspace{-\baselineskip}
\bibliographystyle{mnras} 
\bibliography{refs.bib}

\clearpage

\end{document}